\documentclass[useAMS,usenatbib]{mn2e}
\usepackage{amssymb}
\usepackage{bm}
\usepackage{esint}
\usepackage{graphicx}
\usepackage{epstopdf}
\usepackage[a4paper,pdfstartview=FitV,colorlinks=True,citecolor=blue]{hyperref}

\newcommand{\bea}{\begin{eqnarray}}
\newcommand{\eea}{\end{eqnarray}}
\newcommand{\beq}{\begin{equation}}
\newcommand{\eeq}{\end{equation}}

\title[The nature of voids I]{The nature of voids: I. Watershed void finders and their connection with theoretical models}
\author[S. Nadathur and S. Hotchkiss]{S. Nadathur$^{1}$\thanks{seshadri.nadathur@helsinki.fi} \& S. Hotchkiss$^{2}$\\
$^1$Department of Physics, University of Helsinki and Helsinki Institute of Physics, PO Box 64, FI-00014, University of Helsinki, Finland\\
$^2$Department of Physics and Astronomy, University of Sussex, Falmer, Brighton, BN1 9QH, UK\\
}

\begin{document}

\date{\today}
 
\pagerange{\pageref{firstpage}--\pageref{lastpage}}

\label{firstpage}

\maketitle

\begin{abstract}
The statistical study of voids in the matter distribution promises to be an important tool for precision cosmology, but there are known discrepancies between theoretical models of voids and the voids actually found in large simulations or galaxy surveys. The empirical properties of observed voids are also not well understood. In this paper, we study voids in an $N$-body simulation, using the {\small ZOBOV} watershed algorithm. As in other studies, we use sets of subsampled dark matter particles as tracers to identify voids, but we use the full-resolution simulation output to measure dark matter densities at the identified locations. Voids span a wide range of sizes and densities, but there is a clear trend towards larger voids containing deeper density minima, a trend which is expected for all watershed void finders. We also find that the tracer density at void locations is usually smaller than the true density, and that this relationship depends on the sampling density of tracers. We show that fits given in the literature fail to match the observed density profiles of voids. The average enclosed density contrast within watershed voids varies widely with both the size of the void and the minimum density within it, but is always far from the shell crossing threshold expected from theoretical models. Voids with deeper density minima also show much broader density profiles. We discuss the implications of these results for the excursion set approach to modelling such voids.
\end{abstract}

\maketitle



\section{Introduction}
\label{section:intro}

The study of large underdense voids in the large-scale matter distribution of the Universe has become increasingly important in recent years, with the creation of a number of public catalogues of voids in galaxy survey data \citep{Pan:2012,Sutter:2012wh,Nadathur:2014a} and a wide variety of statistical analyses based on them.

Voids are interesting primarily because of the cosmological information they may contain. Various studies have suggested that they could be used to constrain the expansion history of the Universe and the equation of state of dark energy \citep[e.g.][]{Ryden:1995,Lee:2009,Bos:2012,Hamaus:2014b}, to test modified theories of gravity (\citealt*{Li:2012,Clampitt:2013,Cai:2015}; \citealt{Zivick:2014}), to calibrate measurements of galaxy bias (\citealt{Hamaus:2013}; \citealt*{Chan:2014}), to constrain initial conditions of structure formation \citep*{Kamionkowski:2009}, or to probe more exotic theories such as coupled dark energy \citep{Sutter:2014d}. The primary void observables used in such studies are their abundances and size distributions, the distortion of their shapes in redshift space \citep{Alcock:1979}, their dark matter density profiles, and the void-galaxy or void-void position correlations. Given the exciting potential applications, a rigorous comparison of  theoretical predictions of these properties and those seen for voids in $N$-body simulations and galaxy surveys is very important.

However, this aim is complicated by the degree of ambiguity surrounding a very fundamental question: what exactly \emph{is} a `void'? From a theoretical perspective, there is a clear answer, provided by the spherical evolution model of \citet{Sheth:2003py}, who identify voids as those non-linear underdense regions which have evolved to reach \emph{shell crossing}. This identification is convenient, as voids can then be modelled analogously to collapsed overdense haloes using the excursion set formalism \citep{Press:1974,Bond:1991,Lacey:1993}, and therefore allows clear predictions to be made for void observables.

In practical terms, however, the definition of a void is not so clear. When dealing with either $N$-body simulations or galaxy survey data, an algorithmic approach is required to identify regions as voids, which is complicated by the fact that voids are naturally poorly sampled by observable tracers, making shot noise a serious issue. A number of different void finders have been proposed \citep[see][for a review of methods]{Colberg:2008}, which unfortunately do not always agree with each other. Watershed void finders (e.g., \citealt*{Platen:2007qk,Neyrinck:2008,Sousbie:2011a,Cautun:2013}) form an interesting class of algorithms. They use tessellation techniques \citep{Schaap:2000} to reconstruct the density field from discrete data points, and the watershed algorithm for creating a void hierarchy.  They present a number of advantages for practical studies, as they are more resilient to shot noise in the density reconstruction \citep[though see also e.g.][for other interesting proposals]{Elyiv:2015,Shandarin:2014}, and do not make prior assumptions about void geometries. They are also the most commonly used. Watershed voids may therefore be considered a reasonable practical definition of a void.

However, watershed algorithms make no reference to shell crossing, which is the defining characteric of theoretical models. The obvious question is therefore how, or whether, these two void definitions are related to each other. The answer is important to the practical use of watershed voids in cosmology, as well as to the development of further theoretical predictions. A number of studies (e.g., \citealt{Sutter:2014b,Chan:2014,Pisani:2014,Chongchitnan:2015}; \citealt*{Achitouv:2015}) which apply the \citet{Sheth:2003py} formalism to describe watershed voids \emph{assume} that the two approaches describe the same or closely related objects. There is known to be a significant disagreement between model predictions for void abundances as a function of their size and results obtained for watershed voids in dark matter simulations. This can be partially resolved (at least at large void sizes) by the ad hoc assumption that shell crossing and void formation occur at less extreme densities than predicted by the spherical model, but as we will also show, such an approach lacks self-consistency. A more direct comparison of void properties with the model is therefore desirable.

At the same time, a number of properties of watershed voids remain imperfectly understood. A generic property of watershed void finders is that voids containing the deepest density minima should have the largest sizes. Yet the fits provided by \citet*{Hamaus:2014a} to describe density profiles about void centres appears to suggest the opposite behaviour \citep[though note that the applicability of this fitting form is not universally accepted, e.g.][]{Nadathur:2015a}. Perhaps related to this problem is the question of how to define the `centre' of a void --- which is also important for void correlation studies. The standard procedure assigns the centre to a weighted average of the positions of the tracers of mass within a void \citep[e.g.][]{Lavaux:2012,Sutter:2012wh,Nadathur:2014a,Sutter:2014b}, but it is not clear that this will correctly identify the region with the greatest \emph{absence} of mass. Another interesting question is how densities reconstructed from discrete tracer distributions relate to the true underlying density field. In studies of voids from simulations, the full simulation output is typically randomly down-sampled to provide a set of tracers, which can have a dramatic effect on the recovered void properties \citep{Sutter:2014b}. 

Our goals in this paper are two-fold. We wish to understand the relationship between watershed voids and theoretical models. To do this, we move beyond the fitting of void number functions alone and identify other important characteristics of voids which can be used to test the assumption of shell crossing more broadly. We also want to empirically examine the properties of watershed voids in simulations in order to understand the working of the algorithm and clarify some of the issues above. 

To do so we make use of the popular {\small ZOBOV} watershed algorithm \citep{Neyrinck:2008}. To enable comparison of our results with others in the literature, we will mostly use the options for {\small ZOBOV} implemented in the {\small VIDE} toolkit \citep{VIDE:2015}. We analyse voids identified using randomly subsampled dark matter particles as tracers, and relate them to true densities determined from the full resolution simulation output. We propose a new definition of the void centre, which is designed to better identify the true location of the underdensity within the void. The details of the $N$-body simulation, the watershed algorithm and the methods for identifying void centres and measuring density profiles are described in Section~\ref{section:numerical}.

Section~\ref{section:theory} provides a summary of the spherical model for void evolution, which we use to extract general identifying characteristics of shell-crossed voids for comparison with simulation results. This comparison is performed in Section~\ref{section:properties}, where we also outline the general properties of voids in our simulation. We show that larger voids do correspond to deeper density minima, as expected for the watershed algorithm. We examine the viability of the fitting formula of \citet{Hamaus:2014a} to describe the density profiles of voids, and show that the fits described in that paper do not provide a good quantitative or qualitative description of the variation of the average profile within the void population. In addition, subsampled tracers almost always overestimate the density contrast in voids. All of these results have practical implications for future studies that use watershed void finders. We compare these results from simulated voids to theory and argue that there is no evidence that watershed void finders in general, and {\small VIDE} and {\small ZOBOV} in particular, satisfy the primary defining criteria of the \cite{Sheth:2003py} model. This leads us to reassess the viability of describing watershed voids using existing theoretical techniques. We summarize and conclude in Section~\ref{section:conclusions}.


\section{Numerical methods}
\label{section:numerical}

\subsection{Simulation}
\label{subsec:MultiDark}
In this paper we use $N$-body simulation data from the MultiDark Run1 (MDR1) release of the MultiDark project \citep{Prada:2012}.\footnote{Publicly available from \url{www.cosmosim.org}.} MDR1 uses an Adaptive-Refinement-Tree (ART) code, based on adaptive mesh refinement, to simulate $2048^3$ dark matter particles within a cubic volume of $1\,(h^{-1}\mathrm{Gpc})^3$, in a $\Lambda$CDM cosmological model with parameters $(\Omega_m,\Omega_\Lambda,\Omega_b,h,n_s,\sigma_8)=(0.27,0.73,0.0469,0.7,0.95,0.82)$. The simulation has mass resolution $m_p = 8.721\times10^9\;h^{-1}M_\odot$ and force resolution $7\;h^{-1}$kpc. Initial conditions were set  using the Zeldovich approximation at redshift $z=65$.

From the full particle output at redshift 0 we randomly subsample the dark matter particles down to a number density of $\overline{n}=3.2\times10^{-3}\;h^3$Mpc$^{-3}$, similar to that of typical galaxy samples \citep[e.g.][]{Zehavi:2011}. This corresponds to a mean nearest-neighbour separation of $\overline{n}^{-1/3}\sim7\;h^{-1}$Mpc. We refer to the resulting sample as the \emph{Main} sample, and use these particles as tracers for the void finding. In addition, we have used a control sample with a higher tracer density $2\times10^{-2}\;h^3$Mpc$^{-3}$, which we refer to as the \emph{Dense} sample. However, our primary conclusions regarding the properties of watershed voids do not depend strongly on the tracer number density. Therefore unless otherwise stated, all results presented in this paper refer to voids from the \emph{Main} sample.

Note that a random subsampling of tracers introduces shot noise but does not change the fundamental clustering properties of the dark matter field. Therefore despite having the same tracer number density, the properties of voids in such subsampled tracers and those in the galaxy distribution would not be expected to be (and are not) the same, since galaxies are biased tracers of the matter density. However for our purposes of understanding the general properties of watershed voids in this work, a random subsampling is sufficient. We consider the effects of galaxy bias separately in a companion paper \citep{Nadathur:2015c}. 

Although our tracers are themselves dark matter particles, as the subsampling procedure increases shot noise we will distinguish between the tracer \emph{number} density, and the underlying dark matter density. The dark matter density in MDR1 is determined from the full resolution particle output of the simulation at redshift 0, by using a cloud-in-cell interpolation on a $1024^3$ grid, followed by smoothing with a Gaussian kernel with width equal to one grid cell. The sub-Mpc resolution of this grid is much smaller than the typical void size scales, so that this procedure in effect provides a continuous underlying dark matter density, of which the subsampled tracer particles are an approximately Poisson realization. 

In the following, we will reserve the symbols $\rho$ and $\Delta$ for dark matter densities determined using this gridded smoothed density field, and use the symbols $n$ and $\Delta_n$ for the equivalent quantities determined from the tracer number densities.

\subsection{Void finding}
\label{subsec:voidfinding}

To identify voids in the dark matter particle distribution, we make use of the {\small ZOBOV} watershed void finder \citep{Neyrinck:2008}, with the options implemented in the {\small VIDE} toolkit \citep{VIDE:2015}. Although there are known issues with the application of {\small VIDE} to galaxy survey data with irregular survey volumes and masks \citep{Nadathur:2014a}, when dealing with a simulation cube with periodic boundary conditions these do not present a problem. However, note that in some cases, especially the definition of the void centre described below, we use our own modification of the {\small ZOBOV} algorithm.

{\small ZOBOV} works by reconstructing the density field based on a Voronoi tessellation of the simulation cube around the discrete distribution of tracer particles. The tessellation associates each particle with a Voronoi cell consisting of the region of space closer to it than to any other particle. The volume of the Voronoi cell $i$ relative to the mean volume is then used to estimate the local tracer number density $n_i$ at the particle location. This reconstruction, known as the Voronoi tessellation field estimator (VTFE), is naturally scale-adaptive and thus far more resilient against shot noise effects than naive counts-in-cells measurements.

After reconstructing the density field, the algorithm identifies all local minima of the reconstructed density field and determines the ``catchment basins" around each minimum, known as zones. Zones are then merged to form a nested hierarchy of voids according to the watershed principle \citep{Platen:2007qk}, such that the zone with the smallest minimum density $n_\mathrm{min}$ then acquires  neighbouring higher-density zones as sub-voids, in increasing order of the minimum density on the watershed ridge separating them. For each void thus formed, we define an effective void radius $R_v$ as the radius of a sphere with equivalent volume $V$,
\beq
R_v = \left(\frac{3}{4\pi}V\right)^{1/3}\;.
\eeq

Even in the absence of any merging, deeper density minima typically correspond to zones of larger volume and thus larger $R_v$. However, the watershed merging procedure also ensures that voids with deepest density minima contain greater numbers of merged sub-voids and therefore have the largest sizes. This correlation of minimum density and void size is a common property of all watershed void-finders and is not unique to the {\small ZOBOV} algorithm.

To avoid excessive merging leading to essentially infinite void sizes, {\small VIDE} imposes a restriction preventing the merger of two zones unless the minimum link density along the watershed ridge separating them satisfies $n_\mathrm{link}<n_\mathrm{max}$, where $n_\mathrm{max}=0.2\overline{n}$. This condition applies only to the lowest density point on such a ridge, and does not prevent voids from containing regions of much higher densities. The value of 0.2 therefore has no theoretical motivation, and $n_\mathrm{max}$ should be considered an arbitrary free parameter. Indeed alternative values of $n_\mathrm{max}$ have been considered in other works \citep{Nadathur:2014a,Hotchkiss:2015a,Achitouv:2015,Nadathur:2015a}, and properties such as the abundance of root-level voids, the distribution of void sizes and void density profiles will all depend on the value chosen. A fuller discussion of the effects of this arbitrary choice is provided by \citet{Nadathur:2015c}. However, for ease of comparison with previous results we shall restrict ourselves in this paper to the default value hard-coded in {\small VIDE}.

Further selection cuts might be desirable at this stage, since {\small ZOBOV} simply reports all local density minima as potential voids, without regard to the value of the minimum density within them or any reference to shell crossing. {\small VIDE} provides an optional selection cut which purports to remove those voids which have a tracer number density within a defined central region greater than $0.2$ times the mean. However, this density is measured by naive number counting on a scale smaller than the mean inter-particle separation. Therefore, as pointed out by \citet{NH:2013b}, it is very badly affected by shot noise and the values determined by {\small VIDE} are almost completely uncorrelated with the true central density. In any case this cut only excludes a small fraction of final voids, so we do not apply it. We also do not apply the much more conservative cuts on $n_\mathrm{min}$ suggested by  \citet{Nadathur:2014a}, \citet{Hotchkiss:2015a} and \citet{Nadathur:2015a}. 

A selection cut based on the void radius has sometimes been advocated in the literature, to remove voids with $R_v<\overline{n}^{-1/3}$, which are claimed to be below the resolution limit. In fact the adaptive nature of the tessellation means that {\small ZOBOV} automatically excludes small voids below its resolution limit, as we show in Section \ref{section:properties}. Therefore no further cut on $R_v$ is necessary. 

By applying these criteria we find in total 27 450 voids in the \emph{Main} tracer sample. Of these, 26 919 are root-level voids in the hierarchy, i.e. they are not subvoids of any parent voids and their volumes do not overlap each other. 

\subsection{Void centres}
\label{subsec:centres}
Since voids obtained from the watershed algorithm have arbitrary shapes, different prescriptions may be used to define the location of the void `centre'. The most commonly used definition, which is also the definition implemented in {\small VIDE}, is the volume-weighted barycentre of the void member particles
\beq
\label{eq:barycentre}
\bm{X}^\rmn{bc}_v = \frac{1}{\sum_i V_i}\sum_i V_i\bm{x}_i\;,
\eeq
where $\bm{x}_i$ is the position of the $i$th particle, $V_i$ is the volume of it's corresponding Voronoi cell, and the sum runs over all member particles of void $v$.

In the low-density interior of a void, Voronoi cells in the tessellation are typically greatly elongated, and the particles contained within them lie far from their geometrical centres. This means that the position $\bm{x}_i$ corresponding to each cell is an imprecise measure of the location of the cell. In addition, watershed voids contain a great number of member particles -- the median number for our void sample is $82$, and many voids contain several hundreds -- the vast majority of which reside in the overdense walls and filaments on the outskirts of voids. A combination of these two factors means that although the barycentre position defined by equation~\ref{eq:barycentre} is roughly symmetrically located with respect to the overdense void walls, it is typically very far from the position of minimum density. This is because the barycentre definition is fundamentally based on the locations where tracers are present, rather than  locations from where they are \emph{absent}. A consequence of this is that the location of the barycentre is very sensitive to the sub-void fraction and thus to the arbitrary condition controlling void merging described above.

For some purposes, it may be more logical to define the void centre to coincide with the location of the minimum density within it. This is particularly important when void centre locations are subsequently used to measure density-dependent effects, such as void lensing or ISW contributions. To achieve this, we adopt the following procedure. We identify the core particle of the void as the particle with the largest Voronoi cell (i.e., corresponding to the minimum tracer density $n_\mathrm{min}$), and examine the tessellation output to identify all Voronoi cells adjacent to it. From this set we select the lowest density neighbouring particle, and then, in order of increasing density, two other particles that are adjacent to both the core particle and the previous selections. This provides us with the four lowest density mutually adjacent Voronoi cells in the void; we now define the void centre to lie at the point of intersection of these four cells, which is also the circumcentre of the tetrahedron formed by the four tracer particles. This point represents the location within the void that is maximally distant from all tracers. 

We shall refer to this alternative definition of the void centre as the \emph{circumcentre} and denote its location by $\bm{X}^\rmn{cc}_v$. In Section~\ref{section:properties} we show that both the number density of tracers and the underlying dark matter density are indeed significantly lower at the circumcentre than the barycentre. In Appendix \ref{appendixB} we also show that the circumcentre location is more resilient to shot noise effects arising due to subsampling.

\subsection{Density profile determination}
\label{subsec:profile_methods}
A fundamental quantity of interest is the average distribution of tracers and dark matter about the void centre, and the variation of this distribution with void properties. We study this behaviour by constructing stacked density profiles for subsets of voids satisfying different criteria. To do so we rescale distances within each qualifying void in units of the void radius $R_v$, and then estimate the average density in the stack in concentric spherical shells about the void centre.

Estimating the tracer number density in this way is complicated by shot noise effects, since the interiors of voids by definition contain very few tracer particles which can be used for number density measurements. \citet{Nadathur:2015a} showed that an unbiased estimate accounting for Poisson noise can be obtained using the volume-weighted estimator for the average number density in the $j$th radial shell, 
\beq
\label{eq:Poissonestimator}
\overline{n}^j=\frac{\sum_{v=1}^{N_v}N_i^j+1}{\sum_{i=1}^{N_v}V_i^j}\,,
\eeq
where the $j$th shell has width $\Delta\tilde{r}$ in units of the rescaled radial distance $\tilde{r}$ for each void, $V_i^j$ is the true volume of the $j$th shell of the $i$th void and $N_i^j$ is the number of tracer particles contained within it, and the sum over $i$ runs over all voids included in the stack. Note that the $N_i^j$ in this formula includes all tracer particles within the shell, not just those that are identified as members of the void by the watershed algorithm. Under the assumption that the individual numbers $N_i^j$ are Poisson realizations of the true underlying density, the error in equation~\ref{eq:Poissonestimator} can then be estimated at any desired confidence level directly from the definition of the Poisson distribution. In this paper plotted errorbars indicate the 68\% confidence limits on $n$. 

As the resolution of the dark matter density field is much finer than the typical void size, no such complications are required when estimating $\rho$ over the stack of voids. We simply sample the dark matter density at all grid points contained within the radial shell and calculate the mean and standard deviation of the values obtained. We present our results for the stacked dark matter densities in terms of the average total enclosed density within a radius $r$, $1+\overline{\Delta(r)}$, since this allows a more direct contact with the theory described in Section~\ref{section:theory}.

All density profiles are calculated out to three times the void radius, and are measured in radial bin steps of $0.1$ times the radius.


\section{Excursion set models of voids}
\label{section:theory}

Most existing theoretical descriptions of voids are derived from the framework presented by \citet{Sheth:2003py}. This in turn derives from the original excursion set approach of \citet{Press:1974,Epstein:1983,Bond:1991} and is based on the model of spherical evolution of mass shells \citep{Gunn:1972,Lilje:1991}. In this Section we briefly summarize such models in order to highlight the key areas of comparison with the results of watershed void finders.

In this picture the evolution of a spherical mass shell of radius $r$ is determined by the total enclosed density contrast within the radius of the shell at time $t$, $1+\Delta(r,t)$, where
\beq
\label{eq:Delta}
\Delta(r,t) = \frac{3}{r^3}\int_0^r\left[\frac{\rho(y,t)}{\overline\rho(t)}-1\right]y^2dy,
\eeq
and by the time evolution of the cosmological density parameter $\Omega(t)$. Underdense spherical regions contain a density deficit (i.e., $\Delta(r,t)<0$) which causes shells to expand outwards. This deficit is stronger for inner shells, which therefore expand faster than outer shells, and mass evacuated from the centre of the underdensity begins to pile up at its edges. For a steep enough starting density profile, at some point in the evolution inner shells catch up with shells which were initially further out from them, in an event known as \emph{shell crossing}. The moment of shell crossing marks a transition in the evolution of the underdensity, as it subsequently expands outwards self-similarly \citep*{Suto:1984,Fillmore:1984,Bertschinger:1985}. 

Within the spherical model, it can be shown that shell crossing occurs when the average density enclosed within the void is
\beq
\label{eq:shellcrossing}
\rho_\rmn{enc}/\overline{\rho}=1+\Delta(r,t) \simeq0.2.
\eeq
This corresponds to a linearly extrapolated average density contrast of
\beq
\label{eq:linDelta}
\Delta_\rmn{lin}=\delta_\rmn{v}\simeq-2.71\;,
\eeq
at the epoch of shell crossing, with this value independent of radius $r$ and largely independent of the cosmological parameters governing the background evolution. This is analogous to the case of spherical collapse of clusters, which occurs above a linear overdensity threshold of $\Delta_\rmn{lin}=\delta_\rmn{c}\simeq1.69$.

Following \citet{Blumenthal:1992,Dubinski:1992}, \citet{Sheth:2003py} then identify the population of \emph{voids} with only those mature evolved underdensities that have reached the stage of shell crossing. If the initial Gaussian density fluctuation field is smoothed on a range of different smoothing scales $R$, this physical picture identifies fluctuations which exceed the density threshold $\delta_\rmn{v}$ on smoothing scale $R$ with potential voids of radius $R$ today. These fluctuations can be characterized by their depth in units of the rms fluctuation of the density field on scale $R$,\footnote{Note that the definition $\nu=\delta_\rmn{v}/\sigma_0$ is also commonly used. In this case equation~\ref{eq:SvdWdist} would need to be appropriately modified, as done by \citet{Chan:2014}.}
\beq
\label{eq:nu}
\nu\equiv\delta_\rmn{v}^2/\sigma_0^2(R)\;,
\eeq
where $\sigma_0(R)$ is one of the set of spectral moments
\beq
\label{eq:sigma}
\sigma_j^2(R)\equiv\int\frac{k^{2+2j}}{2\pi^2}W^2(kR)P(k)\rmn{d}k\;,
\eeq
with $P(k)$ the power spectrum of the unsmoothed density fluctuation field and $W(kR)$ the smoothing filter.

At this point, the abundance and size distribution of voids can be predicted by a number of models of varying degrees of sophistication. \citet{Sheth:2003py} use the excursion set approach \citep[e.g.,][]{Bond:1991,Sheth:1998} to account for fluctuations which cross the $\delta_\rmn{v}$ threshold on some small scale but are overdense with $\Delta_\rmn{lin}>\delta_\rmn{c}$ on some larger scale. Such underdensities would be crushed by the collapse of the surrounding cluster and so would not be visible as voids today (the \emph{void-in-cloud} effect). This amounts to a two-barrier problem. According to this model, assuming void number density is conserved on evolving from Lagrangian to Eulerian space, this number density can be expressed as a function of the Eulerian void radius $R_v$ (e.g., \citealt*{Jennings:2013}; \citealt{Chan:2014}) as
\beq
\label{eq:numdens}
\frac{\rmn{d}N}{\rmn{d}R_v} = \left(\frac{3}{4\pi R_\rmn{L}^3}\right) f(\nu)\frac{\rmn{d}\nu}{\rmn{d}R_\rmn{L}}\,,
\eeq
where
\beq
\label{eq:SvdWdist}
f(\nu)\simeq \sqrt{\frac{1}{2\pi\nu}}\exp\left(-\frac{\nu}{2}\right)\exp\left(-\frac{|\delta_\rmn{v}|}{\delta_\rmn{c}}\frac{\mathcal{D}^2}{4\nu}-2\frac{\mathcal{D}^4}{\nu^2}\right)\;,
\eeq
and
\beq
\label{eq:SvdWD}
\mathcal{D}\equiv \frac{|\delta_\rmn{v}|}{(\delta_\rmn{c}+|\delta_\rmn{v}|)}\;.
\eeq
Here the Lagrangian radius $R_\rmn{L}=0.58R_v$, a relationship determined by the shell crossing condition above. 

However, it is well known that equation~\ref{eq:numdens} does not provide a good fit to the distribution of voids found by watershed algorithms, since it predicts a sharp cutoff in void sizes above $\sim5\;h^{-1}$Mpc, much smaller than observed for watershed voids. A number of studies \citep{Jennings:2013,Sutter:2014b,Chan:2014,Pisani:2015}) have attempted to improve fits  by relaxing the shell crossing condition $\delta_\rmn{v}=-2.71$ and treating $\delta_\rmn{v}$ as a free parameter instead.\footnote{\citet{Jennings:2013} also propose an alternative adaptation of this model, but this cuts off the distribution at even smaller $R_v$, so would make the discrepancy worse.} This procedure is not justified by any specific theoretical model. However, if $\delta_\rmn{v}$ is allowed to vary, self-consistency requires that the relationship between Eulerian and Lagrangian radius be correspondingly modified to
\beq
\label{eq:LagrangianR}
R_\rmn{L} = \frac{R_v}{\left(1-\delta_\rmn{v}/c\right)^{c/3}}\,,
\eeq
where $c\simeq1.594$ \citep{Bernardeau:1994,Jennings:2013}. Despite this modification, equation~\ref{eq:numdens} with a variable $\delta_\rmn{v}$ still fails to describe the distribution of small voids, and the fit values of $\delta_\rmn{v}$ for large voids vary widely. \citet{Chan:2014} obtain $\delta_\rmn{v}\simeq-1$ and find little redshift dependence of this value, contrary to theoretical expectation --- however, they keep $R_\rmn{L}=0.58R_v$ fixed when varying $\delta_\rmn{v}$ instead of using equation~\ref{eq:LagrangianR}, so their model is not self-consistent. On the other hand, \cite{Sutter:2014b} find a range of different $\delta_\rmn{v}$ values for voids from different samples, ranging from $-0.26$ to $-0.5$. \cite{Pisani:2015} quote $\delta_\rmn{v}=-0.45$. It is hard to conceive of an explanation for why $\delta_\rmn{v}$ should lie so far from the theoretical prediction if the shell crossing model were true.

\begin{figure}
\includegraphics[width=85mm]{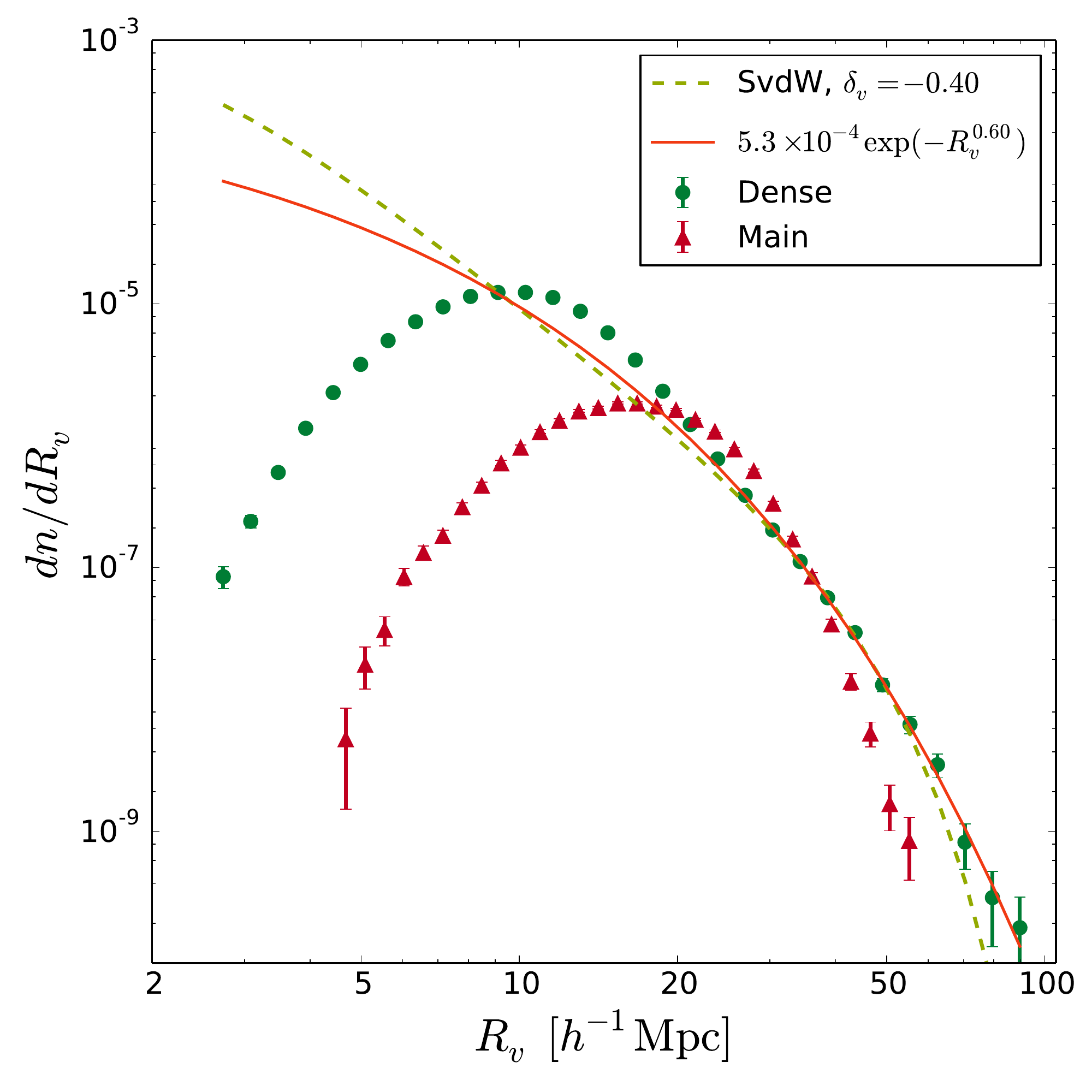}
\caption{The differential number density of voids in simulation as a function of their size, for both simulation samples. Error bars are calculated assuming the void numbers in each bin are Poisson distributed. The dashed line shows the best fit of the \citet{Sheth:2003py} model to the $R_v>25\;h^{-1}$Mpc data, with $\delta_\rmn{v}=-0.40$. The solid line shows an exponential cutoff model which describes the same data better.} 
\label{fig:numdens}
\end{figure}

In any case, insofar as allowing $\delta_\rmn{v}$ to vary allows a fit to the distribution of the largest voids, it does so simply by replicating an exponential cutoff in the distribution at large $R_v$. To demonstrate this, in Fig.~\ref{fig:numdens} we show the distribution of void sizes obtained from both our simulation samples, together with the prediction obtained from eqs.~\ref{eq:numdens} and \ref{eq:LagrangianR} with value $\delta_\rmn{v}=-0.40$, and a simple exponential curve $\propto\exp(-R_v^{0.60})$. These numerical values represent the best fits obtained from fitting to the $R_v>25\;h^{-1}$Mpc data from the higher resolution \emph{Dense} sample (note that the size distribution at large $R_v$ is itself resolution-dependent). The minimum radius cut is imposed since neither model can fit the data at all scales. The best-fit parameters are sensitive to the exact choice of this cut, but the relative quality of the fits does not change significantly. Despite having an extra parameter, the exponential cutoff model significantly outperforms the modified excursion set model on the basis of the Akaike Information Criterion \citep{Akaike:1974}. 
 
An approximate exponential cutoff at large void radii is a rather generic feature of alternative descriptions of voids, and would also apply if for instance voids were modelled simply as minima of a Gaussian density field at a fixed smoothing scale (\citealt{BBKS}; \citealt*{Flender:2013}; \citealt{Nadathur:2014b}). We do not intend to attempt a fully-fledged alternative theoretical description of voids here; instead our point is that the ability or otherwise of equation~\ref{eq:numdens} with variable $\delta_\rmn{v}$ to fit the void distribution in a limited size range should not be taken as evidence that the excursion set model provides a good description of watershed voids. 

However, other aspects of the excursion set model can also be directly tested. The crucial ingredient of the model is identification of underdensities as voids \emph{only if they have undergone shell crossing}. Various modifications of the model (e.g., \citealt*{Paranjape:2012a,Musso:2012,Paranjape:2012b}; \citealt{Jennings:2013,Achitouv:2015}) do not change this fundamental picture. Since watershed algorithms in general make no explicit reference to shell crossing when defining a `void', it is desirable to test this assumption much more directly than through the ad hoc fitting of the number function described above.

A key property of shell crossing is that it occurs at the same enclosed density contrast $\Delta$ for all voids, irrespective of their size. (This is also what justifies the use of a single $\delta_\rmn{v}$ in fitting void abundances.) Adding additional complexities to the spherical evolution model, such as the effect of a shear field, could relax the condition $\Delta=-0.8$ to some extent, and may introduce a small scatter in the values of $\Delta$ over the void population. Nevertheless, strong variation of the enclosed density with properties of watershed voids would be a clear sign that they do not correspond to similar shell-crossed objects.

A related property of shell-crossed voids is that if the enclosed density contrast is to be the same for voids of all sizes, equation~\ref{eq:nu} requires that larger voids must correspond to more extreme fluctuations of the parameter $\nu$. It can be shown that this in turn means that larger voids should on average correspond to shallower but broader initial density profiles $\delta(r)$, while smaller voids correspond to deeper and steeper profiles. That is, smaller shell-crossed voids should contain deeper density minima than large voids.\footnote{We thank Ravi Sheth for drawing our attention to this point.} The analogous situation for collapsing haloes is that the most massive haloes should be the least centrally concentrated, which is indeed the case \citep*[e.g.][]{NFW:1996,NFW:1997}. More generally, voids with the deepest density minima should have the steepest density profiles, and vice versa.

These qualitative properties provide clear tests of the assumption that watershed voids have undergone shell crossing. However, as we show in the next Section, neither of them hold true for the voids obtained using {\small VIDE} and {\small ZOBOV}, nor should one expect them to hold for other watershed void finders.


\section{Properties of watershed voids}
\label{section:properties}

\begin{figure*}
\hspace{4em}\includegraphics[width=130mm]{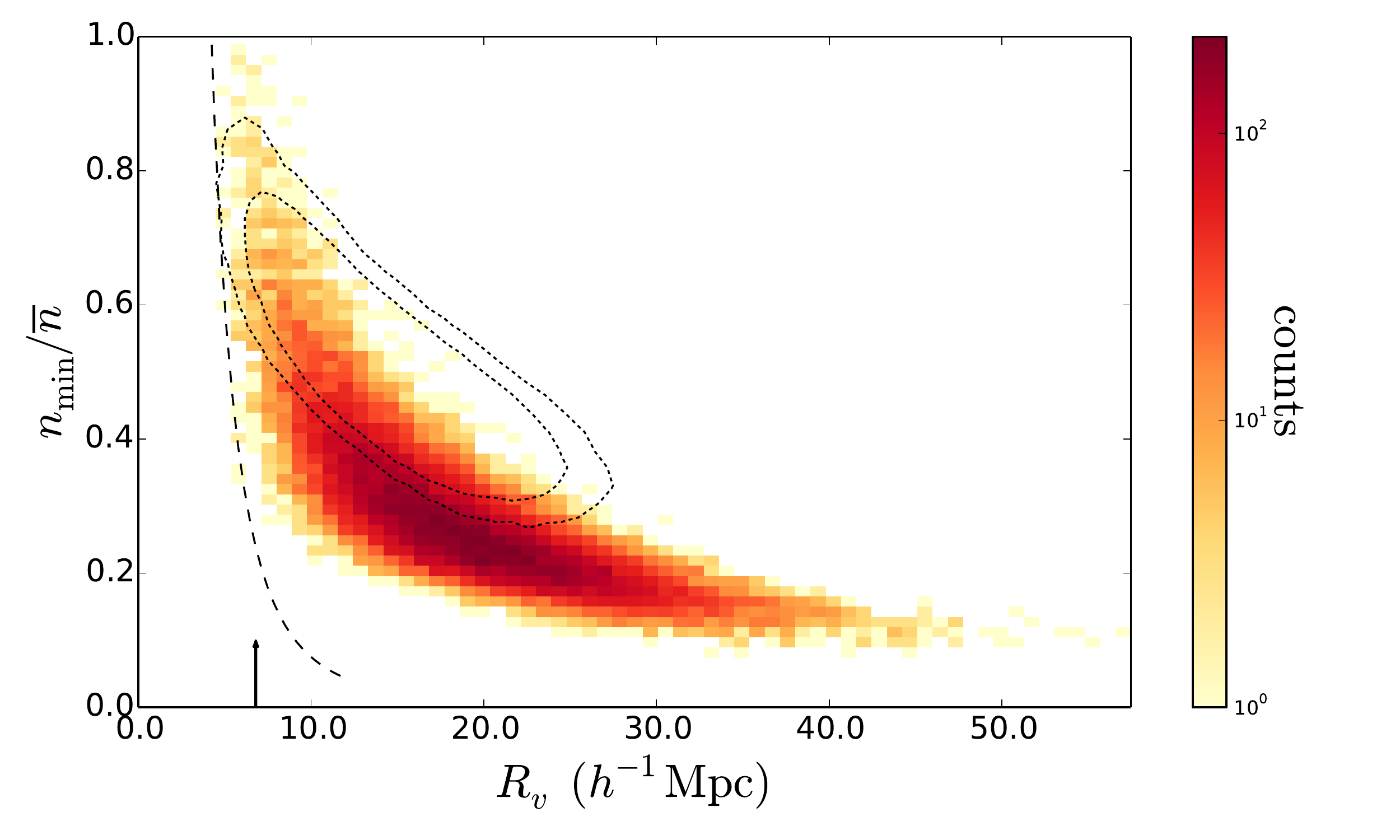}
\caption{The distribution of the minimum tracer number densities within voids and void sizes in the \emph{Main} sample. There is a clear trend towards increasing void size as the minimum density decreases. The dotted lines show the contours enclosing $95\%$ and $99\%$ of all `voids' identified in a random uniform distribution of points with the same number density and in the same volume. The arrow indicates the value $R_v=\overline{n}^{-1/3}$, roughly the mean inter-particle separation, which has sometimes been suggested as minimum size cut. The dashed line shows the true minimum achievable void size resolution as a function $n_\rmn{min}$: most voids automatically lie well away from this limit.} 
 \label{fig:nmin_R}
\end{figure*}

\subsection{Sizes and densities}
\label{subsec:sizes}

Fig.~\ref{fig:nmin_R} shows the distribution of void sizes and minimum tracer number densities for all voids in our \emph{Main} tracer sample. It is immediately obvious that lower minimum number densities are correlated with larger void sizes, as we argued would always be the case for watershed void finders in general and {\small ZOBOV} in particular. The characteristic banana-shaped distribution is similar to that found by \citet{Nadathur:2015a,Nadathur:2015c}, indicating that this is a universal property of the void finder, independent of whether the tracer particles used are dark matter particles or galaxies in haloes. It is also noteworthy that neither {\small VIDE} nor {\small ZOBOV} impose any restriction on the allowed minimum densities, resulting in a range of $n_\rmn{min}$ values extending all the way up to the mean.

The minimum achievable void size resolution is dictated by the process of reconstructing the density field from the Voronoi tessellation described in Section~\ref{subsec:voidfinding}, which sets a natural cutoff of $R_{v,\,\rmn{min}}=\left(3/4\pi\right)^{1/3}r_{N}\left(n_\rmn{min}/\overline{n}\right)^{-1/3}$, where $r_{N}\equiv\overline{n}^{-1/3}\sim7\;h^{-1}$Mpc is roughly the mean inter-particle separation. This cutoff is shown by the dashed black line, and $r_N$ by the vertical arrow. In fact most voids naturally lie well away from this limit, simply because most zones contain several tracer particles. This means that the selection criterion $R_v>r_N$ advocated by some studies \citep[e.g.][]{Sutter:2012wh,Sutter:2014b} has no practical effect, whereas a tighter criterion $R_v>2r_N$ \citep{Hamaus:2014a} is unnecessarily conservative. 

Also shown are contours showing the $95$ and $99$ per cent confidence limit contours for the distribution of spurious `voids' identified by the same algorithm in a random uniform distribution of points with the same volume and same number density as the \emph{Main} sample. There is clearly a considerable overlap between the two distributions, but care is required in its interpretation, as $P\left((n_\rmn{min},R_v)|\rmn{Poisson}\right)$ is not the same as $P\left(\rmn{Poisson}|(n_\rmn{min},R_v)\right)$. In the absence of information on the true dark matter content of such voids, conservative cuts to the void catalogue based on the properties of Poisson voids have previously been advocated \citep[e.g.][]{Neyrinck:2008,Nadathur:2014a, Hotchkiss:2015a}, but these may use available data sub-optimally. Since we have access to this information from the simulation, we analyse all voids without imposing such cuts a priori, and in fact we find that while Poisson contamination increases in the overlap region, statistically speaking voids of all $(n_\rmn{min},R_v)$ values on average correspond to true dark matter underdensities. Similarly, we find that low values of the density ratio $r$ \citep{Neyrinck:2008} also do not serve as a reliable indicator of Poisson contamination. This issue is discussed further in Appendix \ref{appendixB}.

\begin{figure}
\includegraphics[width=85mm]{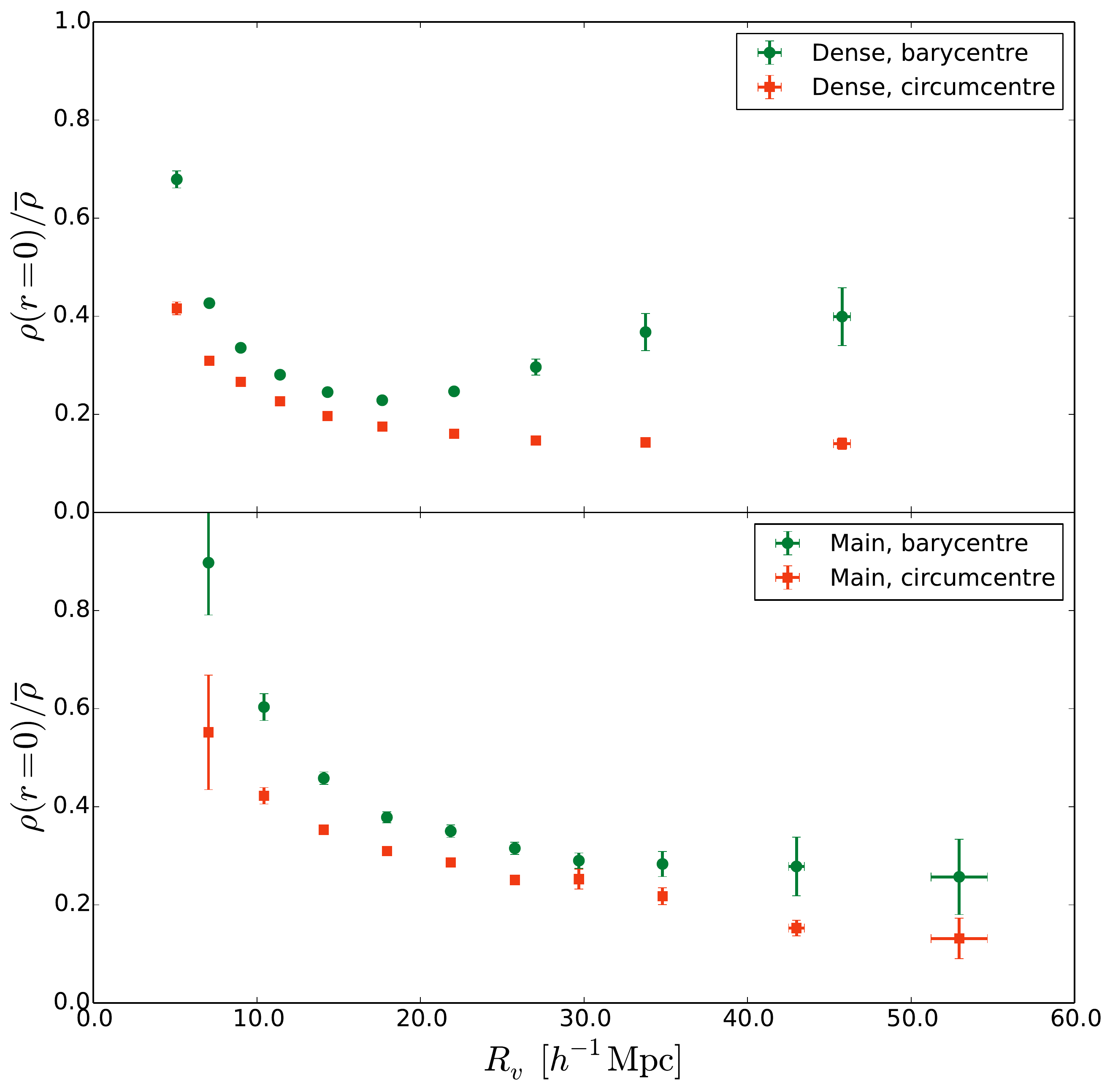}
\caption{Binned average values of the dark matter density at the location of the void centre, as a function of the void size. The top and bottom panels show data for voids in the \emph{Dense} and \emph{Main} samples respectively. Circles (green) and squares (red) refer to the two alternative definitions of the void centre; the circumcentre is clearly a better locator of the true minimum density in the void.} 
\label{fig:rho_R}
\end{figure}

\begin{figure*}
\hspace{4em}\includegraphics[width=125mm]{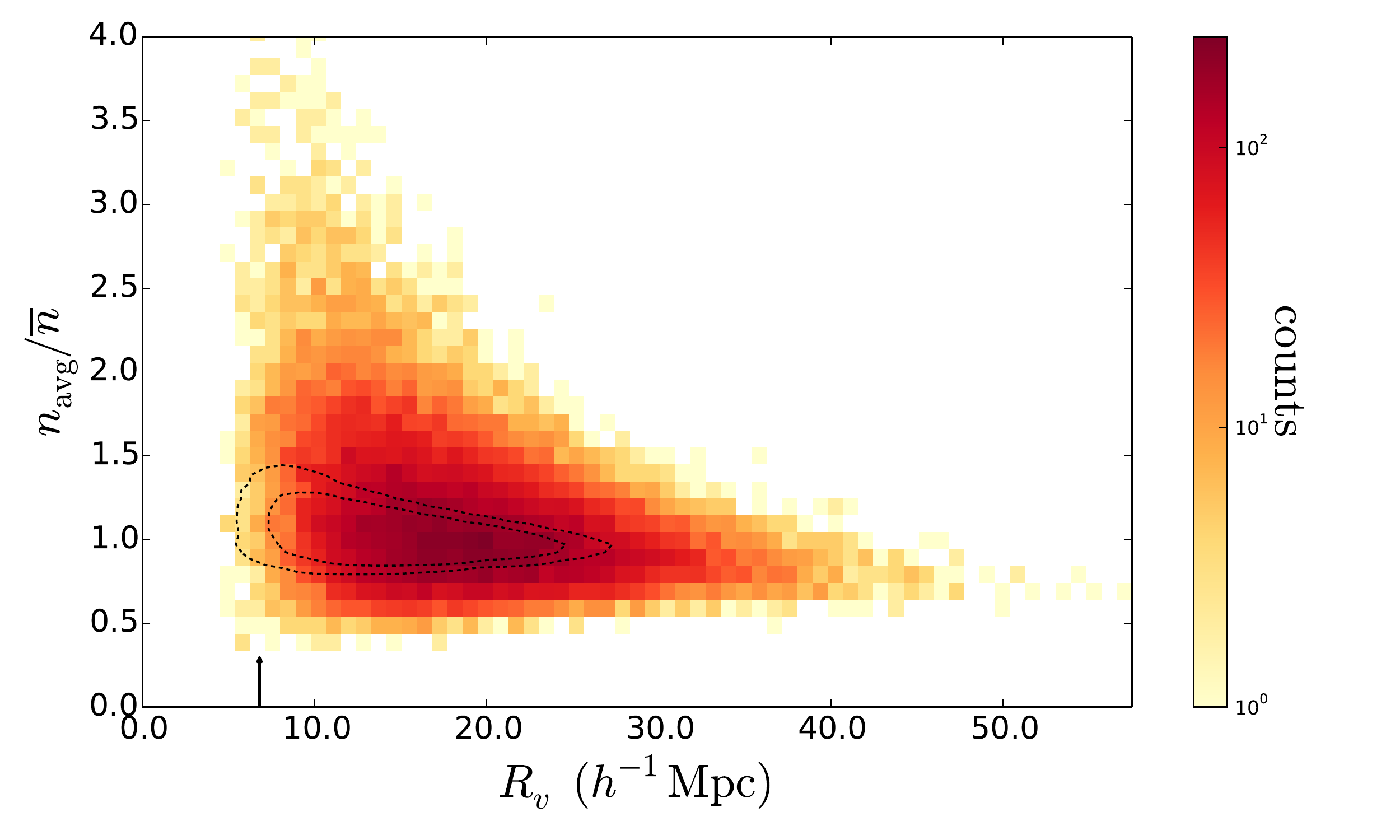}
\caption{The distribution of the average tracer number densities within voids and void sizes in the \emph{Main} sample. The dotted lines show the contours enclosing $95\%$ and $99\%$ of voids in random distributions as in Fig.~\ref{fig:nmin_R}, and the arrow indicates the approximate mean inter-particle separation.} 
 \label{fig:navg_R}
\end{figure*}

Another noteworthy aspect of Fig.~\ref{fig:nmin_R} is the apparent saturation of the minimum densities within voids with increasing $R_v$. This is a consequence of the finite tracer number density, and the saturation value is dependent on the mean density $\overline{n}$. Subsampling tracer particles lowers $\overline{n}$ and thus reduces the apparent tracer density contrast in voids. Conversely, \citet{Nadathur:2015c} show that at the same mean tracer density, more highly biased tracers result in much lower values of $n_\rmn{min}$ within voids.

Given the uncertainties associated with the tracer number density discussed below, the true dark matter density at void locations is perhaps a more informative quantity. To measure this we make use of the dark matter density field described in \ref{subsec:MultiDark} and simply measure its value $\rho$ in the grid cell corresponding to the position of the void centre. 
Fig.~\ref{fig:rho_R} shows the binned average central densities as a function of the void radius $R_v$, for both \emph{Dense} and \emph{Main} voids, and for both definitions of the void centre described in Section~\ref{subsec:centres}. The error bars represent the $2\sigma$ uncertainty in the mean. As expected, in all cases the circumcentre definition is a superior indicator of the location of minimum density within the void. It is also clear that this minimum density decreases with increasing void size, contrary to the excursion set prediction. Curiously, for voids in the \emph{Dense} sample, the density at the barycentre first decreases and then increases with $R_v$. This is because in this case voids with large $R_v$ tend to be formed from the merger of several sub-voids. As such sub-voids necessarily correspond to shallower density minima they do not affect the location of the minimum density, but they do shift the location of the barycentre via equation~\ref{eq:barycentre}. In contrast, the location of the circumcentre is independent of the sub-void fraction and the choice of criteria to control void merging.

Fig.~\ref{fig:navg_R} shows the distribution of the \emph{average} tracer density $n_\rmn{avg}$ within the void, calculated from the number of void member particles and the void volume, and $R_v$. As pointed out by \citet{Nadathur:2014a,Achitouv:2015}, $n_\rmn{avg}$ is typically $\gtrsim1$ and much larger than $n_\rmn{min}$, simply because the watershed definition means that voids always extend to include high density regions on the separating ridges. This feature of {\small ZOBOV} and {\small VIDE} indirectly demonstrates that most tracers in identified voids reside in overdensities, which explains why the barycentre is a poor locator of the minimum underdensity, and also suggests that when $n_\rmn{avg}>1$ the void radius is a significant overestimate of the size of the true underdense region.

\begin{figure}
\includegraphics[width=85mm]{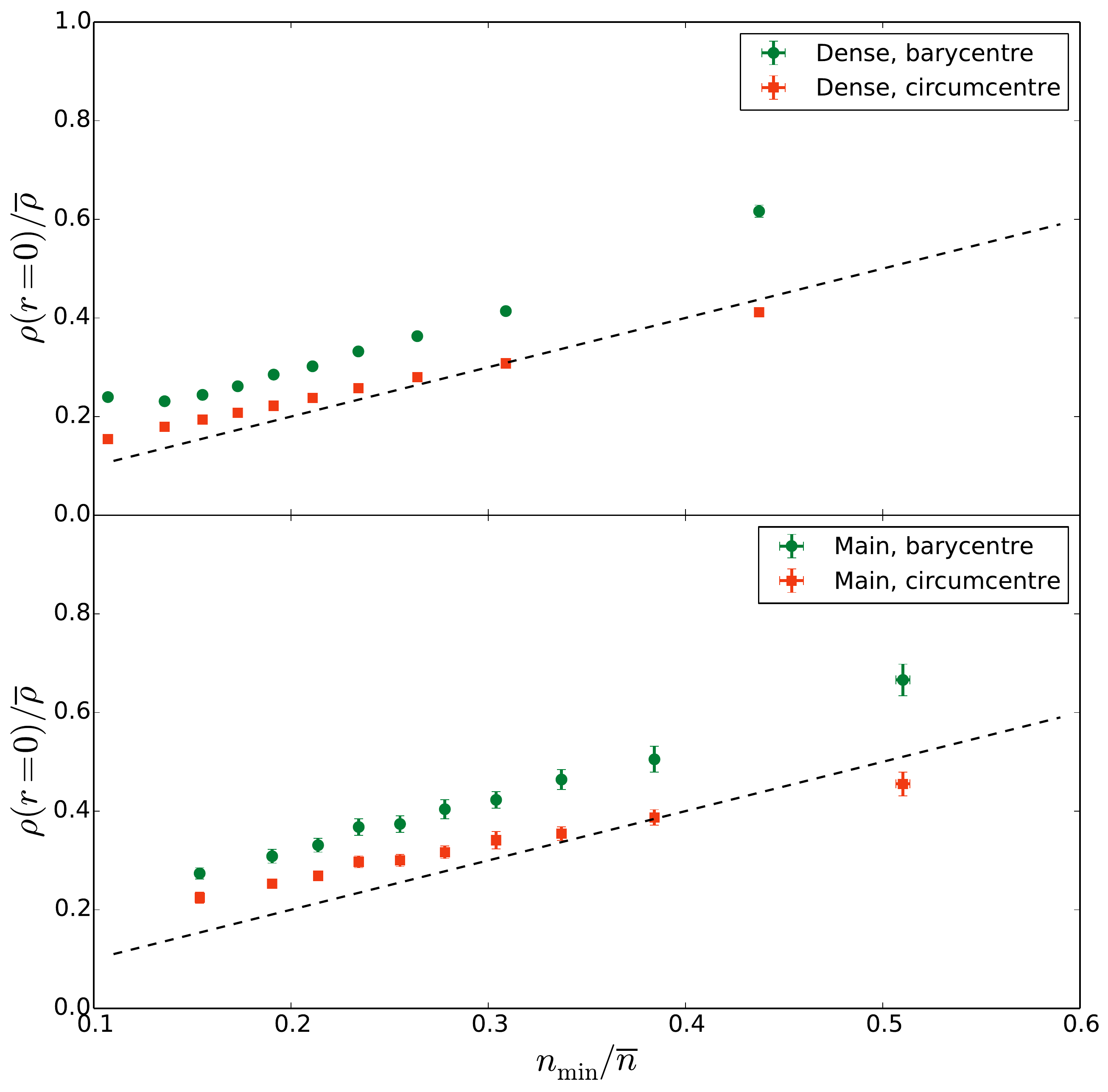}
\caption{Binned average values of the true dark matter density at the location of the void centre, as a function of the minimum tracer number density within the void, for both definitions of the void centre. The top and bottom panels show the data for voids from the \emph{Dense} and \emph{Main} tracer samples respectively. A $45^\circ$ line is shown for reference in each case.} 
\label{fig:rho_nmin}
\end{figure}

It is worth noting that a selection cut on the minimum void radius alone is a sub-optimal way of excluding voids that are on average overdense, since it would eliminate many with the lowest $n_\rmn{avg}$ values as well. 

\subsection{Tracer density versus dark matter density}
\label{subsec:tracervDM}

The relationship between the tracer number density and the true underlying dark matter density in the simulation is also of interest. Even though the tracers in our case are down-sampled dark matter particles, we find that these two quantities are in general not the same. There is already an inherent shot noise in the number densities of dark matter particles arising from the fact that they constitute a discrete realization of the underlying continuous density field. Randomly down-sampling the dark matter particles enhances this shot noise, meaning that, particularly in void regions, tracer number densities tend to be larger than the true dark matter density. This problem is to a large extent mitigated by the self-adaptive nature of the Voronoi tessellation, but cannot be completely removed. A second consequence of reducing the total number of tracers is to increase the effective smoothing scale at which tracer number densities are measured. As mentioned above, for the VTFE density reconstruction this scale is $\sim r_N\sim7\,h^{-1}$Mpc, whereas the dark matter density field is smoothed at a scale of $<1\,h^{-1}$ Mpc. This change of smoothing scales acts in the opposite direction, tending to make dark matter underdensities appear shallower in $n_\rmn{min}$.

The relative strengths of these two effects are illustrated in Fig.~\ref{fig:rho_nmin}, which shows the relationship between the dark matter density $\rho$ at the position of the void centre and the minimum tracer number density within the void as determined from the tessellation, for both definitions of the void centre, and for voids from both the \emph{Main} and \emph{Dense} tracer samples. That $\rho$ exceeds $n_\rmn{min}$ at the barycentre for both samples is to be expected, since the barycentre typically lies quite far from the location of the tracer density minimum. But even at the circumcentre, which is guaranteed to lie in the region of minimum tracer number density, the tracer number density does not accurately reflect the dark matter density, although there is a clear linear relationship between $\rho$ and $n_\rmn{min}$. Particularly in the most underdense voids, shot noise enhancement dominates, causing $n_\rmn{min}/\overline{n}<\rho_\rmn{min}/\overline{\rho}$. However, for the shallowest voids the smoothing effect becomes more important, reversing the relationship. The same qualitative trends are seen for both the \emph{Dense} and \emph{Main} tracer samples, and more generally hold for any tracer population obtained from subsampling dark matter particles.

\subsection{Density profiles}
\label{subsec:profile_results}

\begin{figure*}
\begin{center}
\includegraphics[width=158mm]{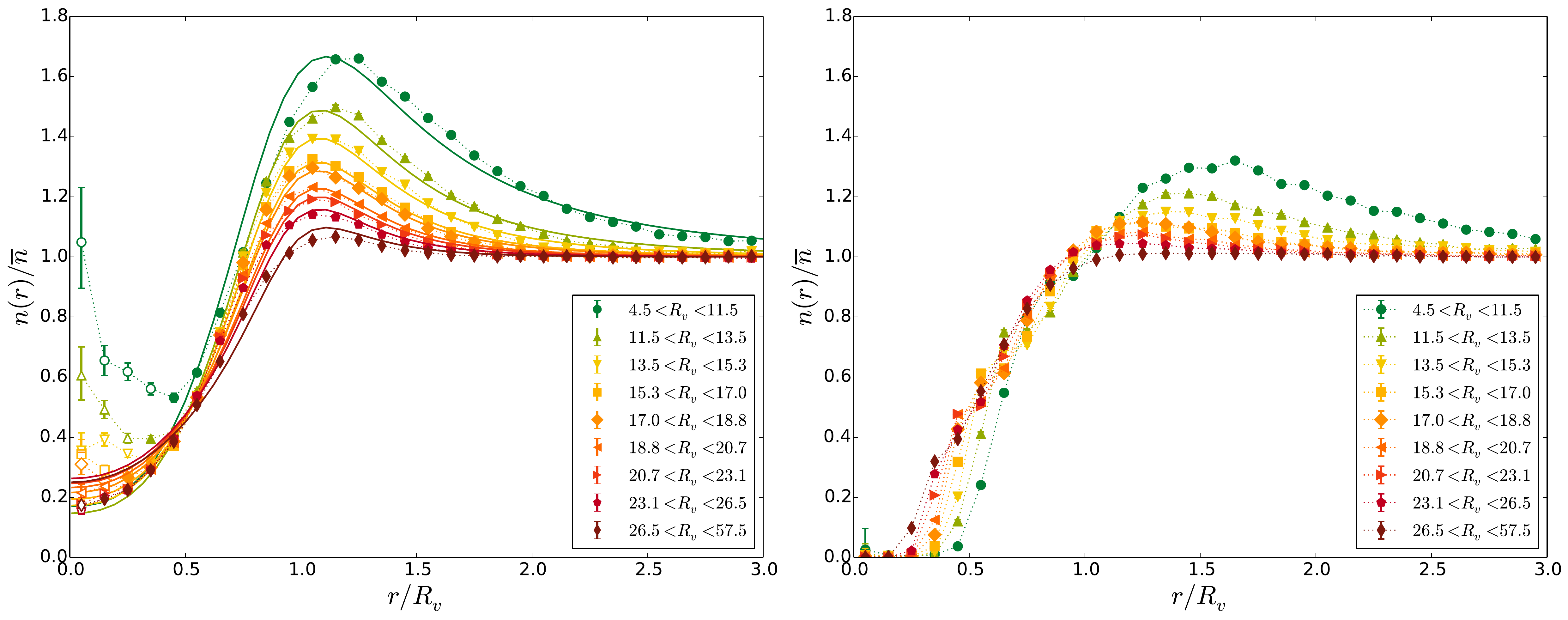}
\caption{Stacked tracer number density profiles for voids of different sizes. Stacks are chosen to include equal numbers of voids in each. \emph{Left}: Profiles for void stacks centred on void barycentres. The solid lines show the best-fit forms of the fitting formula of \citet{Hamaus:2014a,Sutter:2014b} (eqs.~\ref{eq:HSW}, \ref{eq:HSWalpha} and \ref{eq:HSWbeta}), which generally provides a poor fit to the data. Due to discreteness artefacts at small $r$, data points shown with open symbols are excluded from the fitting procedure. \emph{Right}: Profiles for the same voids but with the stacks centred around the circumcentres. By construction the circumcentre more accurately locates the region of minimum tracer density within the void.} 
 \label{fig:n_profiles}
\end{center}
\end{figure*}

We now turn to the distribution of tracer particles and dark matter around void centres. Anticipating that the form of the density profiles will depend on the void size, we first examine the average profiles for stacks of voids within different ranges of $R_v$, chosen such that each stack contains an equal number of voids. 

Fig.~\ref{fig:n_profiles} shows the resulting tracer number density profiles for stacks centred on the void barycentres and circumcentres in the left and right panel respectively. The barycentre stacks show a strong trend for decreasing central density as the void size increases, as is expected from Fig.~\ref{fig:nmin_R}. Voids are generally surrounded by overdense walls, which are much higher for small voids than for large ones. The general asymmetry of the circumcentre location with respect to particles in the void walls is also apparent in the fact that the stacked profiles about this location are less able to resolve the high densities in these walls. On the other hand, central densities are much lower for the circumcentre stacks. This is the essential tradeoff between the two centre definitions: the barycentre has a greater degree of symmetry with respect to the surrounding overdensities, whereas the circumcentre identifies the true location of the \emph{under}density.

A curious feature is apparent in the stacked barycentre profiles at smaller void radii: the tracer density does not show a minimum at the void barycentre, but instead at some distance \emph{away} from the centre. In fact for the smallest voids the average tracer density at the barycentre is indistinguishable from the mean. As argued in Section~\ref{subsec:centres}, the void barycentre is always displaced away from the location of the minimum tracer density within the void; in particular, for small voids the barycentre is often at or very close to the location of a tracer particle within the void. The tracer number density $n(r)$ is measured by naively counting the numbers of tracer particles within volumes on scales generally much smaller than the mean inter-particle separation. As a result, when the barycentre location is close to a tracer particle, high central values for $n(r)$ are obtained.

The converse effect can be seen by comparison with the right panel of Fig.~\ref{fig:n_profiles}, which shows the stacked density profiles for the same voids, but based around the void circumcentre. The circumcentre is also a special point, as it is by construction as far as possible from all tracers in the void. Unsurprisingly therefore, sufficiently small spheres around the circumcentre contain no particles at all and $n(r)\sim0$ for voids of all sizes, even though $n_\rmn{min}$ values are never so small and vary with void size. The Voronoi tessellation avoids this issue because of its self-adaptive resolution. It can be seen from Fig. \ref{fig:rho_nmin} that the Voronoi reconstructed $n_\rmn{min}$ is a much better predictor of the true dark matter density than number counts --- which is why it is preferable to reconstruct the density field from the tessellation in the first place. For this reason, such stacked number density profiles should not be relied upon for quantitative analysis without calibration. 

\begin{figure*}
\begin{center}
\includegraphics[width=158mm]{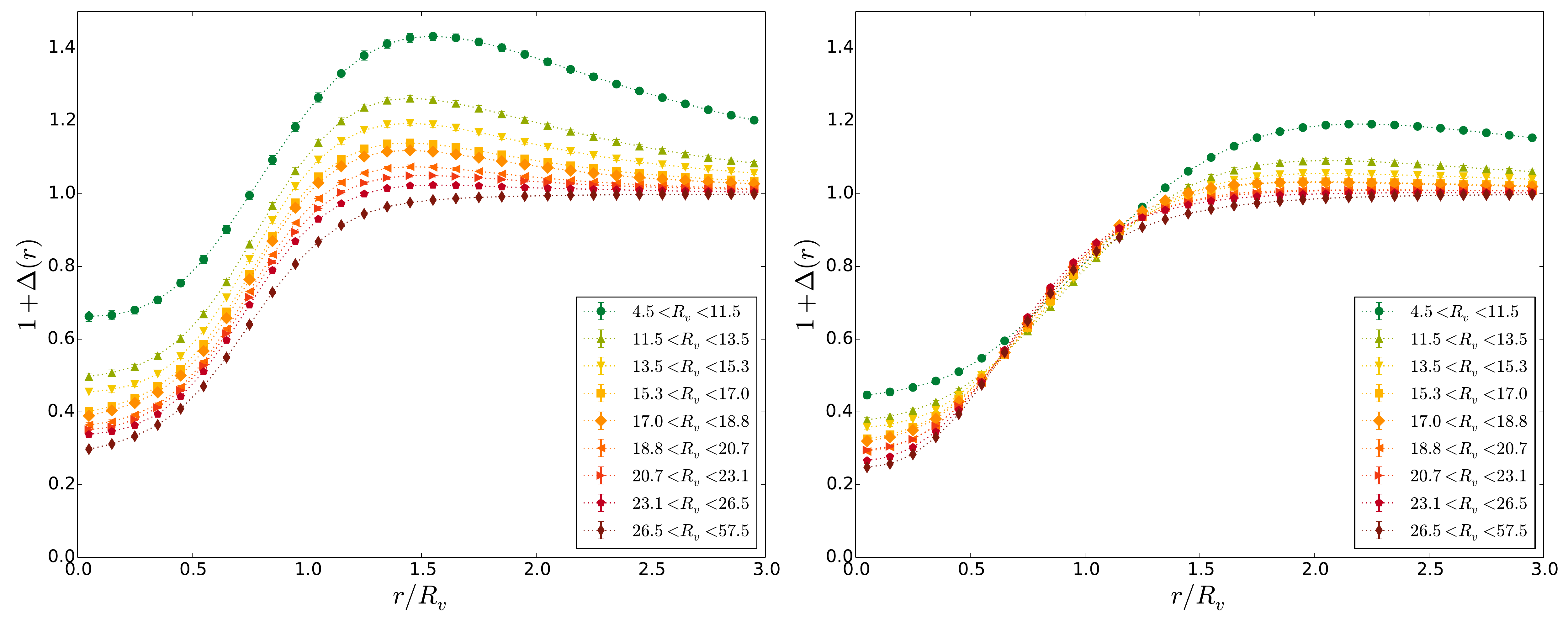}
\caption{Stacked profiles of the total enclosed dark matter density, $1+\Delta(r)$, within radius $r$ of the void centre. The stacks are the same as in Fig.~\ref{fig:n_profiles}. The left panel shows profiles for stacks centred around the void barycentres, and the right panel for stacks centred around the circumcentres.} 
 \label{fig:rho_R_profiles}
\end{center}
\end{figure*}

For this purpose we instead make use of the full dark matter density field at high resolution. Profiles of the average enclosed dark matter density $1+\Delta(r)$ are shown in Fig.~\ref{fig:rho_R_profiles}, for the same void stacks as before. These confirm some properties of watershed voids which are of significance for the attempts to model them theoretically. First, as already seen in Figs.~\ref{fig:nmin_R} and \ref{fig:rho_R}, larger voids contain deeper density minima. Secondly, the enclosed density contrast within these voids is $\Delta(r)>-0.8$, for all void sizes and at all distances $r$. The condition for shell crossing to occur is thus not satisfied at any point within the average void. Finally, the central matter densities are much lower for circumcentre stacks than those centred on the barycentre, as expected from Fig.~\ref{fig:rho_R}.

\begin{figure*}
\begin{center}
\includegraphics[width=160mm]{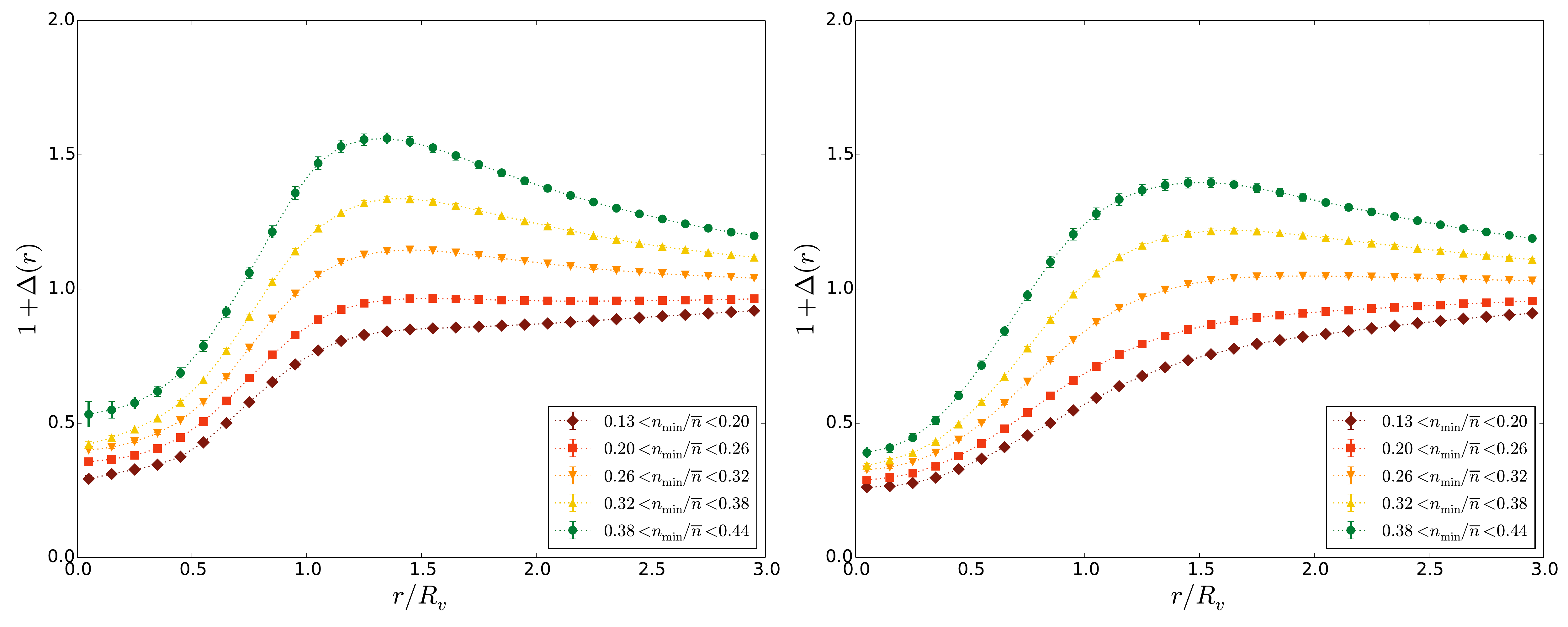}
\caption{Stacked profiles of the total enclosed dark matter density, $1+\Delta(r)$, for voids within the same size range $15<R_v<20\;h^{-1}$Mpc, but with different minimum tracer densities. Profiles in the left panel are stacked about the barycentres, and in the right panel about the circumcentres.} 
 \label{fig:rho_nmin_profiles}
\end{center}
\end{figure*}

Our results in Fig.~\ref{fig:n_profiles} may also be compared to those of \citet{Hamaus:2014a}, who postulate a `universal' profile for voids based on the functional form
\beq
\label{eq:HSW}
\frac{n(r)}{\overline{n}}=1+\delta_c\left(\frac{1-\left(r/r_s\right)^\alpha}{1+\left(r/R_v\right)^\beta}\right)\,.
\eeq
This profile form has two free parameters, $\delta_c$ and $r_s$, with $\alpha(r_s)$ and $\beta(r_s)$ fixed by eqs. \ref{eq:HSWalpha} and \ref{eq:HSWbeta}. Note that although \citet{Hamaus:2014a} refer to this as the density profile $\rho(r)$, their fits to data are in fact based on measurements of the tracer \emph{number} density $n(r)$ --- as emphasized above, for sub-sampled dark matter tracers these two quantities are not the same. We determine the best-fit values of these two parameters by fitting to the $n(r)$ data for each void stack. To avoid the discreteness artefacts described above, in each stack we exclude data points with $\left(r/R_v\right)\overline{R_v}<\left(3/4\pi\right)^{1/3}r_N$ from the fitting procedure, where $\overline{R_v}$ is the mean void radius for the stack. The resulting fits are shown by the solid lines in the left-hand panel of Fig. \ref{fig:n_profiles}.  It can be seen that this `universal' profile function generally provides a poor fit to the data, both within the void interior and in the overdense walls.

Similar results are obtained when fitting to the true dark matter density profiles $\rho(r)$. The behaviour of the best-fit parameters in equation~\ref{eq:HSW} as functions of the void size $R_v$ also significantly differs from that claimed by \citet{Hamaus:2014a} and \citet{Sutter:2014b}. In particular, these authors suggest that the central density contrast $\delta_c$ is an approximately linear, increasing function of $R_v$ for voids of all sizes and in all tracer populations. This behaviour is central to their claim that equation~\ref{eq:HSW} can provide a `universal' description of void density profiles in all tracer populations through a simple rescaling of void sizes. On the contrary, as shown in Fig.~\ref{fig:rho_R}, $\delta_c(R_v)$ is a strictly decreasing function for \emph{Main} sample voids, and is a non-linear, U-shaped function for \emph{Dense} sample voids. Further discussion of the fitting profile is provided in Appendix \ref{appendixA}, where we show that the behaviour of other parameter fits also disagrees with the claimed universality.

We conclude that the fits provided by \citet{Hamaus:2014a,Sutter:2014b} fail both qualitatively and quantitatively to describe the density profiles we observe. Equally, we find no evidence for the self-similarity of tracer density profiles seen by \citet{Nadathur:2015a}, but in this case differences in methodology, the use of dark matter particles instead of galaxies as tracers and the different selection criteria applied to voids may preclude direct comparison (see \citealt{Nadathur:2015c} for a fuller discussion).

So far, following earlier works \citep{Hamaus:2014a,Nadathur:2015a} we have only considered the variation in the mean profile with the size of the voids included in the stack, but it is clear that this cannot be the only important variable. In fact, as shown in Fig.~\ref{fig:rho_nmin_profiles}, voids of similar sizes but different minimum densities $n_\rmn{min}$ have very different density profiles. Voids with different $n_\rmn{min}$ clearly do not enclose the same density contrasts, and deeper density minima do not correspond to steeper density profiles. The enclosed density contrast $\Delta$ clearly varies widely over the void population, providing further evidence that the population of watershed voids does not satisfy the foundational assumption of the excursion set model.

It is also clear that a more complete description of the density profiles around voids is obtained by accounting for the extent of variation in both dimensions of the $\left(n_\rmn{min},R_v\right)$ plane. Fitting formulae such as those provided by \citet{Hamaus:2014a} or \citet{Nadathur:2015a}, which account only for variation with void radius, will in principle be unable to describe the full variety of watershed voids. 

However, it is worth stressing that the distribution of highly biased galaxies trace dark matter underdensities rather differently than the randomly down-sampled dark matter particles we have used in this work \citep{Nadathur:2015c}, and it is the dark matter profiles of galaxy voids which are of greater practical interest in cosmology. 

\section{Conclusions}
\label{section:conclusions}

Our aim in this paper was to provide an empirical investigation into the properties of watershed voids in order to better understand the operation of void finding algorithms such as {\small VIDE} and {\small ZOBOV} and the relation to theory. Several previous studies have focused on the distribution of void sizes alone, and have attempted to fit this using modifications of the spherical evolution model. Such an approach however misses the important relationship between void size and density: \emph{larger voids correspond to deeper density minima}. This is a fundamental feature of {\small ZOBOV} that holds irrespective of whether the tracers used for void identification are simulation dark matter particles, haloes or galaxies. It is also a more general property that should apply to \emph{any} watershed void finder.

The conclusion that follows from this relationship --- and which we also demonstrate directly through stacked density profiles around void centres --- is that  watershed voids cannot correspond to a population of objects which all enclose the same density contrast, which is the principal starting assumption of theoretical descriptions deriving from the model of \citet{Sheth:2003py}. It has long been known that the void number function prediction of this model fails to match that of watershed voids by many orders of magnitude. It has sometimes been argued without proof \citep[e.g.][]{Sutter:2014b,Chan:2014} that the void formation threshold $\delta_\rmn{v}$ might differ from the shell crossing value in the spherical model due to the generally aspherical nature of watershed voids. This assumption has led several authors to treat $\delta_\rmn{v}$ as a free parameter but without altering the basic model. 
However, given the range in enclosed density contrasts $\Delta$ over the watershed void population, a single value of $\delta_\rmn{v}$ for all voids does not seem tenable. Even more suggestive is the fact that for no subset of these voids does the average enclosed density contrast satisfy the criterion for shell crossing, $\Delta\simeq-0.8$, at \emph{any} radial distance from the centre, let alone at the void radius $R_v$. Nor do smaller voids correspond to deeper density minima as expected in the model.

The simplest interpretation of this evidence is that watershed voids simply do not correspond to objects that have undergone shell crossing. With hindsight this should not seem surprising --- {\small ZOBOV} uses only information on the local topology of the density field, and makes no reference to shell crossing. Furthermore, neither {\small VIDE} nor {\small ZOBOV} apply any meaningful conditions even on the minimum tracer density $n_\rmn{min}$ within voids, instead reporting \emph{all} local density minima. Attempts to explain how the shell crossing density criterion may be altered in such voids therefore seem to be misguided. A simpler starting proposition would be to give up the enforced assumption of shell crossing and to describe watershed voids simply as what they are: regions of density minima.

We should stress that breaking this link to theoretical models of shell-crossed voids does not necessarily make the results obtained from watershed void finders less useful for practical cosmological studies. For instance, these voids can still be used to identify large-scale underdense environments. Some of them (though not all) will also correspond to maxima of the gravitational potential, and so they can still be used for studies of lensing \citep{Melchior:2014} or the ISW effect \citep{Cai:2013ik,Hotchkiss:2015a,Planck:2015ISW}. Equally, we do not intend to claim that the \citet{Sheth:2003py} model does not correctly describe shell-crossed underdensities on much smaller scales \citep[although see][]{Falck:2015,Achitouv:2015}. Our statement is simply that this and related models do not match simulation or observational data because the word `void' has a different meaning in the two contexts.

Another interesting feature of our results is the relationship between the underdensity in voids measured using subsampled tracers and using the full resolution dark matter density. We show that values of $n$ and $\rho$ do not completely agree, and apparent tracer underdensities in deep voids are deeper than the true dark matter minima. The relationship between $n$ and $\rho$ depends on the mean sampling density of tracers; it will certainly also change if biased tracers are used. This does not affect the basic operation of {\small ZOBOV}, which only uses relative tracer densities to identify minima, but it argues against the use of absolute values of the central tracer density in applying selection cuts, as has sometimes been suggested \citep[e.g.][]{Sutter:2012wh,Jennings:2013,Sutter:2014b,Nadathur:2014a}. In other words, selecting a region which apparently satisfies the shell-crossing criteria in terms of the tracer number density does not ensure that it does so in the true matter density. This was already pointed out by \citet{Furlanetto:2006} for the case when the tracers are galaxies; our results show that it applies even if the tracers are a subset of dark matter particles in the simulation.

\section{Acknowledgements}

We thank Ravi Sheth for stimulating correspondence and Alexis Finoguenov for helpful discussions. SH acknowledges support from the Science and Technology Facilities Council [grant number ST/L000652/1].

The MultiDark Database used in this paper and the web application providing online access to it were constructed as part of the activities of the German Astrophysical Virtual Observatory as result of a collaboration between the Leibniz-Institute for Astrophysics Potsdam (AIP) and the Spanish MultiDark Consolider Project CSD2009-00064. The MultiDark simulations were run on the NASA's Pleiades supercomputer at the NASA Ames Research Center.

\bibliography{refs.bib}
\bibliographystyle{mn2e}

\appendix
\section{Void density profiles}
\label{appendixA}

In this section we provide a quantitative discussion of the differences between our results on the void density profiles and those of \citet{Hamaus:2014a,Sutter:2014b}. For this purpose we consider only the average profiles for stacks centred on the void barycentres $\bm{X}_v^\rmn{bc}$, and we also select the stacks on the basis of void size $R_v$ alone.  Note that this last step ignores the very strong systematic dependence of profiles on $n_\rmn{min}$ seen in Fig.~\ref{fig:rho_nmin_profiles} and by \citet{Nadathur:2015c}, so is intended solely in order to enable a direct comparison with the earlier results.

We consider both stacked tracer number density and matter density profiles for our \emph{Main} sample voids, which we will characterize in terms of the general fitting function of \citet{Hamaus:2014a},
\beq
\label{eq:genHSW}
\frac{d(r)}{\overline{d}}=1+\delta_c\left(\frac{1-\left(r/r_s\right)^\alpha}{1+\left(r/R_v\right)^\beta}\right)\,,
\eeq
where $d=n,\,\rho$ respectively, and the values of the parameters $\delta_c$, $r_s$, $\alpha$ and $\beta$ are fit to the data in each case. Note that \citet{Hamaus:2014a,Sutter:2014b} only provide fits to $n(r)$, but refer to this as the true density profile. The form of equation \ref{eq:genHSW} considered in these works has only two free parameters, $\delta_c$ and $r_s$, since $\alpha$ and $\beta$ satisfy
\beq
\label{eq:HSWalpha}
\alpha(r_s) \simeq -2.0(r_s/R_v)+4.0\,
\eeq
and
\bea
\label{eq:HSWbeta}
\beta(r_s) = \left\{
\begin{array}{cr}
  \displaystyle 17.5(r_s/R_v)-6.5  & \rmn{for}\;r_s/R_v<0.91, \\
  \displaystyle  -9.8(r_s/R_v)+18.4 &  \rmn{for}\;r_s/R_v>0.91.
\end{array}
\right.
\eea
In Section~\ref{subsec:profile_results} we noted that this restricted two-parameter form does not in general provide a good fit to the $n(r)$ profiles of \emph{Main} sample voids. The same is also generally true for fits to $\rho(r)$. We therefore treat $\alpha$ and $\beta$ as free parameters and compare their best-fit values thus obtained to those predicted by eqs.~\ref{eq:HSWalpha} and \ref{eq:HSWbeta}.

\begin{figure}
\includegraphics[width=80mm]{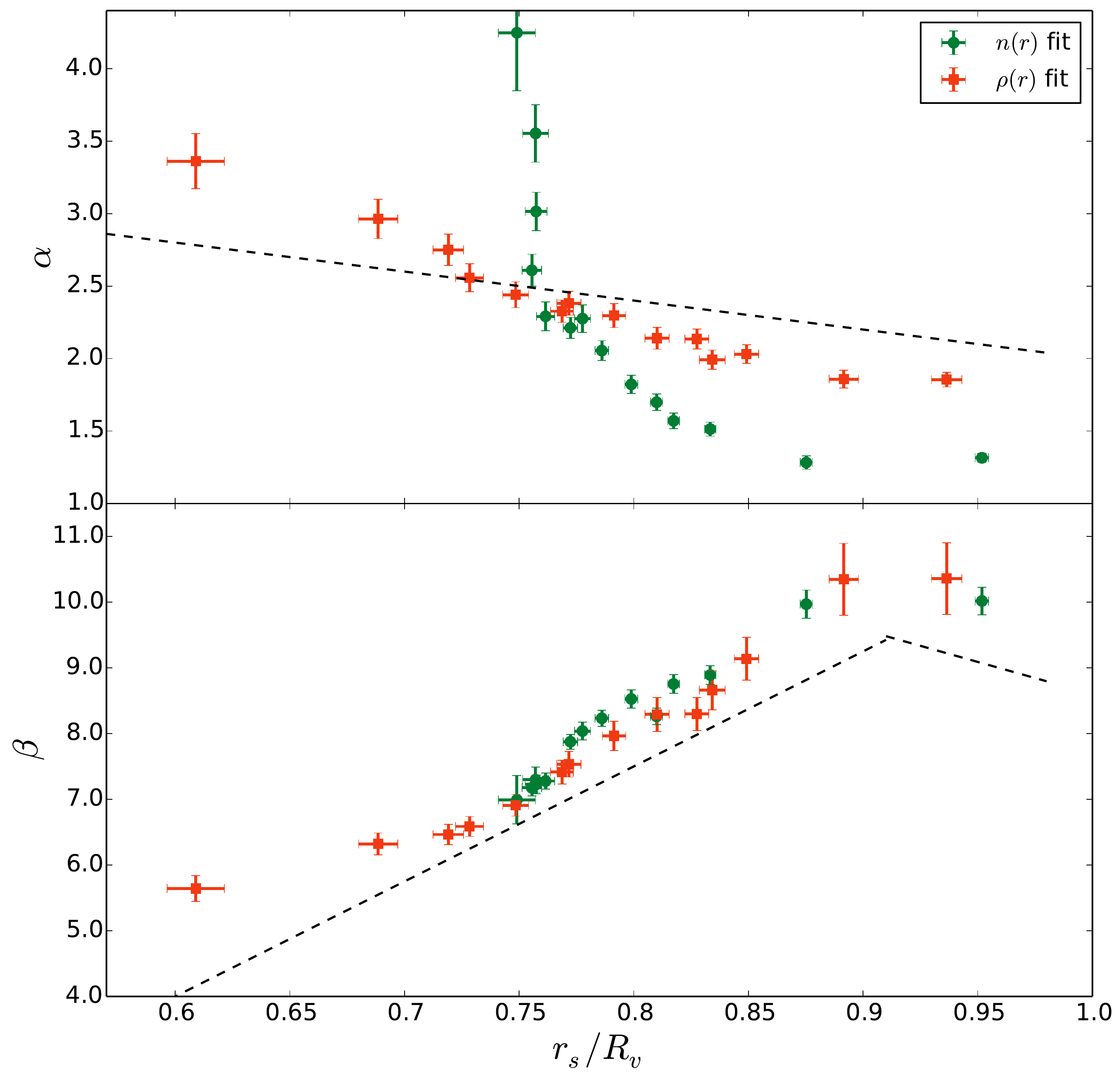}
\caption{The dependence of the best-fit parameters $\alpha$ and $\beta$ in equation~\ref{eq:genHSW} on the value of $r_s$, for fits to the measured tracer number density profiles $n(r)$ (green circles) and true matter density profiles $\rho(r)$ (red squares) for stacks of \emph{Main} sample voids centred about the void barycentres. Stacks are created by binning the voids according the $R_v$ and bins are chosen to contain equal numbers of voids. The black dashed lines represent the expectations if eqs.~\ref{eq:HSWalpha} and \ref{eq:HSWbeta} were correct. } 
\label{fig:HSW2}
\end{figure}

Fig.~\ref{fig:HSW2} shows the best-fit values of $\alpha$ and $\beta$ as functions of $r_s/R_v$ for stacks of different mean void radius $\overline{R_v}$. For comparison, the black dashed lines in each case show the dependence expected from eqs.~\ref{eq:HSWalpha} and \ref{eq:HSWbeta} respectively. The stacks are created by binning the void population on the basis of the void size; the bins are chosen in order to have equal numbers of voids and thus equal statistical power to constrain the profile. Fits are shown for both the tracer number density (green circles) and matter density (red squares) profiles, using the appropriate version of equation~\ref{eq:genHSW} in each case. For tracer number density profiles $n(r)$, data points with $\left(r/R_v\right)\overline{R_v}<\left(3/4\pi\right)^{1/3}r_N$ are not considered in obtaining the fits, due to the artefacts discussed in Section \ref{subsec:profile_results}. Fits to the $\rho(r)$ and $n(r)$ profiles differ significantly from each other, as we have repeatedly emphasized, despite the tracers being a subset of dark matter particles. Both $\alpha$ and $\beta$ are also clearly discrepant with eqs.~\ref{eq:HSWalpha} and \ref{eq:HSWbeta}, meaning that the two-parameter fitting formula of \citet{Hamaus:2014a,Sutter:2014b} is not a good description of the observed void density profiles.

\begin{figure}
\includegraphics[width=80mm]{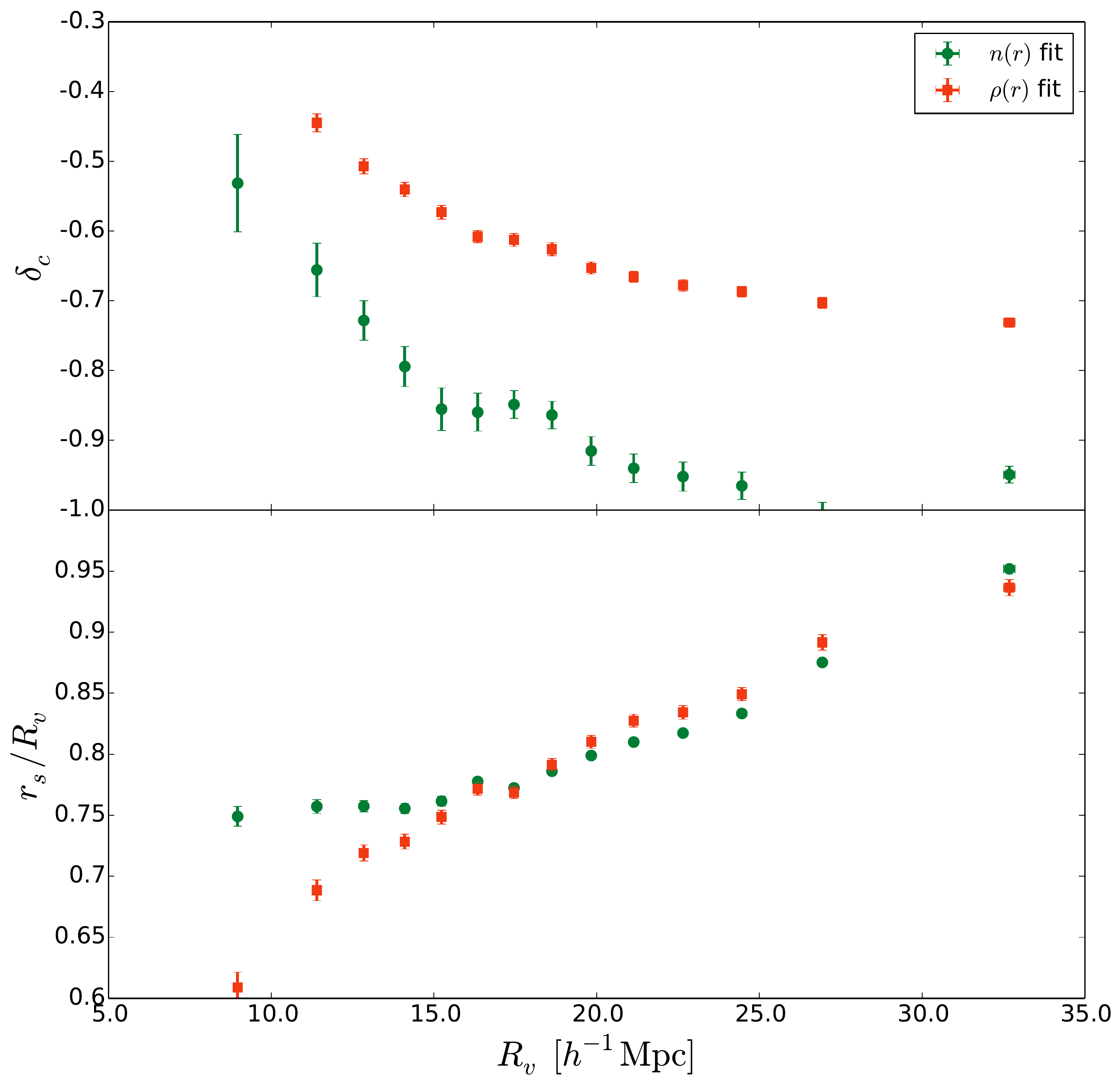}
\caption{The dependence of the best-fit parameters $\delta_c$ and $r_s$ in equation~\ref{eq:genHSW} on the stack mean void radius $R_v$, for the same stacked data as in Fig.~\ref{fig:HSW2}.} 
\label{fig:HSW1}
\end{figure}

Fig.~\ref{fig:HSW1} shows the dependence of the fit values of central density contrast $\delta_c$ and scale radius $r_s$ on the stack mean void radius $\overline{R_v}$ for our voids. Unsurprisingly $n(r)$ and $\rho(r)$ fits again give very different results. Nevertheless, both sets of data points show a clear \emph{decrease} in $\delta_c$ with increasing $R_v$. The same trend is also seen for voids in galaxy populations \citep{Nadathur:2015c}. This is in clear disagreement with the results of \citet{Hamaus:2014a} and \citet{Sutter:2014b}, who suggest $\delta_c(R_v)$ is an increasing function for voids in all tracer populations. 

\citet{Sutter:2014b} also make the stronger claim of universality: that the fitted values of $\delta_c(R_v)$ for voids in any tracer population can be translated to the corresponding values for voids in any other tracer population, irrespective of the sampling density and tracer type, by a simple rescaling of the void radius. Such a rescaling is however only possible if $\delta_c(R_v)$ is approximately linear, and shows the same behaviour (either increasing or decreasing with $R_v$) for voids in all samples.  Fig.~\ref{fig:rho_R} shows that in the \emph{Dense} sample, the density at the void barycentre is far from a linear function of $R_v$, with a clear minimum at $R_v\sim15\;h^{-1}$Mpc. Comparing this to Fig. \ref{fig:HSW1} shows that it is not possible to relate void profiles from these two samples by a rescaling of the void radius. We conclude that the void density profile is \emph{not} universal, but depends at least on the mean density of the tracer population. In \citet{Nadathur:2015c} we argue that it also depends on the tracer bias, and the choice of arbitrary input parameters in the watershed algorithm.

\section{Effects of tracer shot noise on voids}
\label{appendixB}

\begin{figure}
\includegraphics[width=85mm]{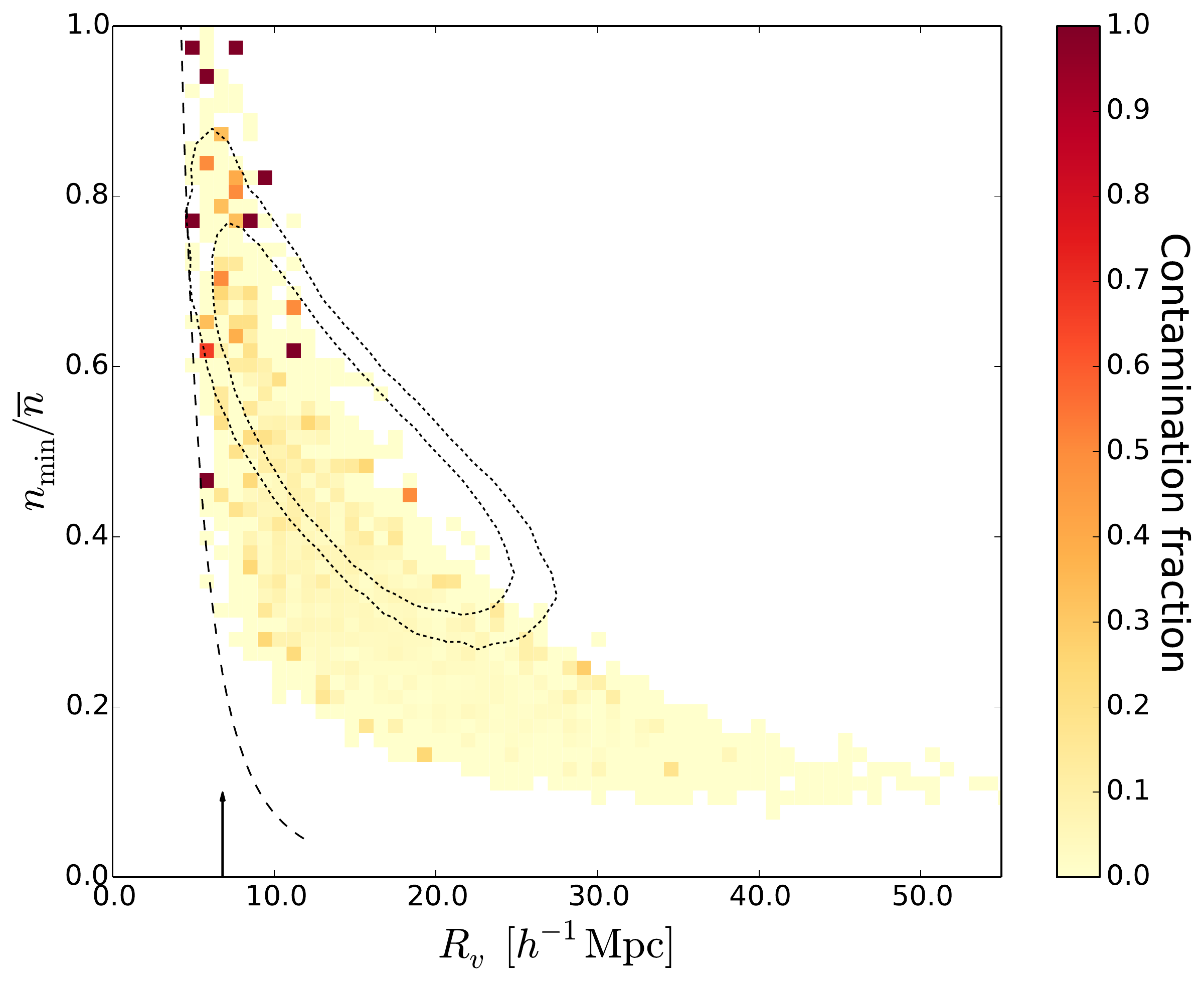}
\caption{The contamination fraction of voids in the \emph{Main} sample. Voids are regarded as genuine if the matter density at their circumcentre locations is less than average, and spurious if not. The Poisson likelihood contours, minimum size resolution and mean inter-particle separation are represented as in Fig.~\ref{fig:nmin_R}.}
\label{fig:Poisson}
\end{figure}

Voids are by definition regions containing a smaller than average number of tracer particles, so their identification is necessarily affected by shot noise problems. To a very large extent, the use of the VTFE reconstruction in the void-finding algorithm mitigates this problem compared to naive methods of estimating tracer number densities. Nevertheless, there is a real possibility that some fraction of the voids returned by the algorithm are spurious detections that are not related to true matter underdensities, especially as {\small ZOBOV} detects large numbers of `voids' even in completely random Poisson point distributions \citep{Neyrinck:2008}. This issue is particularly important when using sub-sampled dark matter tracers.

To quantify the level of possible Poisson contamination we examine the true matter density in the simulation output at the location of the void circumcentre $\bm{X}_v^\rmn{cc}$, which is the best estimate of the location of minimum density. Voids are then classified as `spurious' if they are overdense at this location, $\rho_{cc}/\overline{\rho}>1$, or `genuine' if $\rho_{cc}/\overline{\rho}<1$. This classification deliberately avoids more stringent canonical limits on $\rho_{cc}$ based on comparison with the excursion set model, since this has been shown not to apply to watershed voids in any case. It also uses the circumcentre rather than the barycentre, since the barycentre has been shown to be a worse locator of the true matter underdensity.

Based on this classification, we calculate the contamination fraction. This is shown in Fig.~\ref{fig:Poisson} for the \emph{Main} sample as a function of void observables $n_\rmn{min}$ and $R_v$. The contamination understandably increases slightly at larger $n_\rmn{min}$. There is also possibly a small increase in the contamination fraction within the overlap region with the Poisson contours. However, contamination is generally low and for almost all $(n_\rmn{min},R_v)$ values the majority of voids are genuine. It is also clear that 
 $P\left(\rmn{Poisson}|(n_\rmn{min},R_v)\right)$ is significantly smaller than  $P\left((n_\rmn{min},R_v)|\rmn{Poisson}\right)$ within much of the overlap region. Comparison of the contamination fractions for the \emph{Dense} and \emph{Main} samples shows that, as expected, the contamination fraction increases with subsampling of the tracer population. \citet{Nadathur:2015c} find that at the same sampling density contamination decreases with increased tracer bias.

Shot noise due to subsampling of tracers also reduces the number of detected voids at almost all $R_v$, as shown in Fig.~\ref{fig:numdens}. For the voids that are detected, it can also cause an offset in the recovered locations of the void centre. To investigate the relative stability of the alternative centre definitions $\bm{X}_v^\rmn{cc}$ and $\bm{X}_v^\rmn{bc}$, we matched each void centre in the \emph{Main} sample to its nearest neighbour in the \emph{Dense} sample and examined the distribution of matched nearest neighbour distances in both cases. Subsampling can lead to a significant offset: for the circumcentre the median of this distribution was $0.44R_v$ and the mean $0.51R_v$. The corresponding values for the matched barycentre locations were $7$ and $3$ per cent larger respectively, indicating that the circumcentre definition is somewhat more robust to shot noise effects. We also repeat the matching for voids in different realizations of the same random subsampling: in this case the recovered voids can correspond to rather different subsets of the original voids, so close matches are less common for both centre definitions. However the median matched distance for circumcentres remained smaller than or equal to that for barycentres in all our tests.

\label{lastpage}
\end{document}